\documentclass[a4paper,twocolumn,floatfix,amsmath,amssymb,nobibnotes,altaffilletter]{revtex4}

\usepackage{graphicx}
\usepackage{pslatex}
\usepackage{xspace}
\usepackage{subfigure}

\newcommand{\pedot}{poly(3,4-ethylenedioxythiophene):poly(styrenesulfonate)~}
\newcommand{\degree}{$^{\circ}$}
\newcommand{\pos}{$V_{POS}$}
\newcommand{\voc}{$V_{OC}$}
\newcommand{\vcv}{$V_{CV}$}
\newcommand{\qfb}{$V_{QFB}$}
\newcommand{\vbi}{$V_{Bi}$}
\newcommand{\cgeo}{$C_{geo}$}

\begin{document}

\title{Built-in potential and validity of Mott--Schottky analysis in organic bulk heterojunction solar cells}

\author{M.~Mingebach}\email{markus.mingebach@physik.uni-wuerzburg.de}
\affiliation{Experimental Physics VI, Julius-Maximilians-University of W{\"u}rzburg, D-97074 W{\"u}rzburg}
\author{C.~Deibel}\email{deibel@physik.uni-wuerzburg.de}
\affiliation{Experimental Physics VI, Julius-Maximilians-University of W{\"u}rzburg, D-97074 W{\"u}rzburg}
\author{V.~Dyakonov}
\affiliation{Experimental Physics VI, Julius-Maximilians-University of W{\"u}rzburg, D-97074 W{\"u}rzburg}
\affiliation{Functional Materials for Energy Technology, Bavarian Center for Applied Energy Research (ZAE Bayern), D-97074 W{\"u}rzburg}

\date{\today}

\begin{abstract}

We investigated poly(3-hexylthiophene-2,5-diyl):[6,6]-phenyl-C$_{61}$ butyric acid methyl ester bulk heterojunction (BHJ) solar cells by means of pulsed photocurrent, temperature dependent current--voltage and capacitance--voltage measurements. We show that a direct transfer of Mott--Schottky (MS) analysis from inorganic devices to organic BHJ solar cells is not generally appropriate to determine the built-in potential, since the resulting potential depends on the active layer thickness. Pulsed photocurrent measurements enabled us to directly study the case of quasi flat bands (QFB) in the bulk of the solar cell. It is well below the built-in potential and differs by diffusion-induced band-bending at the contacts. In contrast to MS analysis the corresponding potential is independent on the active layer thickness and therefore a better measure for flat band conditions in the bulk of a BHJ solar cell as compared to MS analysis.

\end{abstract}

%\pacs{71.23.An, 72.80.Le, 73.20.At}

%\keywords{organic semiconductors; polymers; photovoltaic effect; pulsed photocurrent; built-in potential; capacitance--voltage; impedance; band-bending}

\maketitle

%\section{Introduction}
 
Recent progress in organic solar cell development yielded greatly improved power conversion efficiencies up to 8.3\% \cite{Green2011} and even 9.2\% are reported by press media. \cite{Service2011} Nevertheless, further improvements of efficiency and device lifetime are needed and hence the investigation of basic working principles and limitations of organic solar cells is of high interest. \cite{Deibel2010, Brabec2010} One of these important key parameters of organic bulk heterojunction (BHJ) solar cells is the built-in (\vbi) potential, since it influences the internal electric field profile in the devices and also gives an upper limit for the open circuit voltage \cite{Rauh2011} (and hence solar cell efficiency). Therefore the correct determination of the built-in potential is essential to better understand the potential of a given material combination used as the active layer of organic solar cells.\cite{Scharber2006}

Many well established measurement methods have been directly transferred from inorganic devices to organic solar cells with only slight adjustments to the interpretation of the results. This approach may not be sufficient in all cases. For instance, De Vries et al.\ already showed that well established methods such as electroabsorption measurements of organic light emitting diodes do not lead to \vbi~as expected. \cite{DeVries2010}  

Mott--Schottky (MS) analysis is a very well-known and powerful tool to determine the built-in potential \vbi---the difference of the electrode work functions \cite{Malliaras1998}---and the doping concentration of a device, too. Although MS theory is based on the properties of an abrupt pn-junction \cite{Sze2007} it has already been applied to organic devices. \cite{garciabelmonte2008, bisquert2008, nolasco2010}. While the interpretation of the slope of $1/C^{2}(V)$ has been adapted as a concentration of occupied trapping centers (acceptor doping density in the active layer), \cite{Ray2010, bisquert2008} the x-axis intercept is usually identified as the built-in voltage of organic semiconductor devices, \cite{Boix2009} even though the obtained values are lower than expected from the difference of the electrode work functions. Hence we will denote the potential obtained by MS analysis for organic semiconductors as \vcv.
 
We show in this paper that MS analysis is not appropriate to determine the built-in potential of organic bulk heterojunction solar cells. The obtained values differed up to 0.35 V from the built-in potential determined by temperature dependent measurements of the open circuit voltage. \cite{Rauh2011} Furthermore we observed a thickness dependence of \vcv~that contradicts the interpretation of \vcv~as \vbi. In contrast, the case of quasi flat bands (QFB) determined by pulsed photocurrent measurements \cite{limpinsel2010} takes band-bending in vicinity of the contacts into account and showed no dependence on the active layer thickness. Hence it is a better measure for flat band conditions inside a working BHJ solar cell as compared to MS analysis. 
 
%\section{Experimental Methods}

All devices mentioned here were prepared by spin casting a photoactive layer of P3HT:PC$_{61}$BM (1:1 weight ratio in chlorobenzene) onto \pedot (PEDOT:PSS) covered indium tin oxide (ITO) glass substrates. Afterwards the devices were annealed on a hotplate for 10 minutes at 130 \degree C. As final step metal contacts consisting of Ca (3 nm) and Al (100 nm) were applied by thermal evaporation. All processes were performed in a nitrogen atmosphere glovebox. The photovoltaic performance was determined by a Xe arc lamp that was adjusted to simulate standard testing conditions. \cite{Shrotriya2006} All investigated solar cells had active areas of 3 mm$^{2}$ and exhibited power conversion efficiencies of 3-3.5\% and fill factors of 65-70\%.

Afterwards the devices were transferred to a microprobe station (Janis Research ST-500, liquid Nitrogen) in which all temperature dependent measurements were carried out.

\begin{figure}
	\includegraphics[width=7cm]{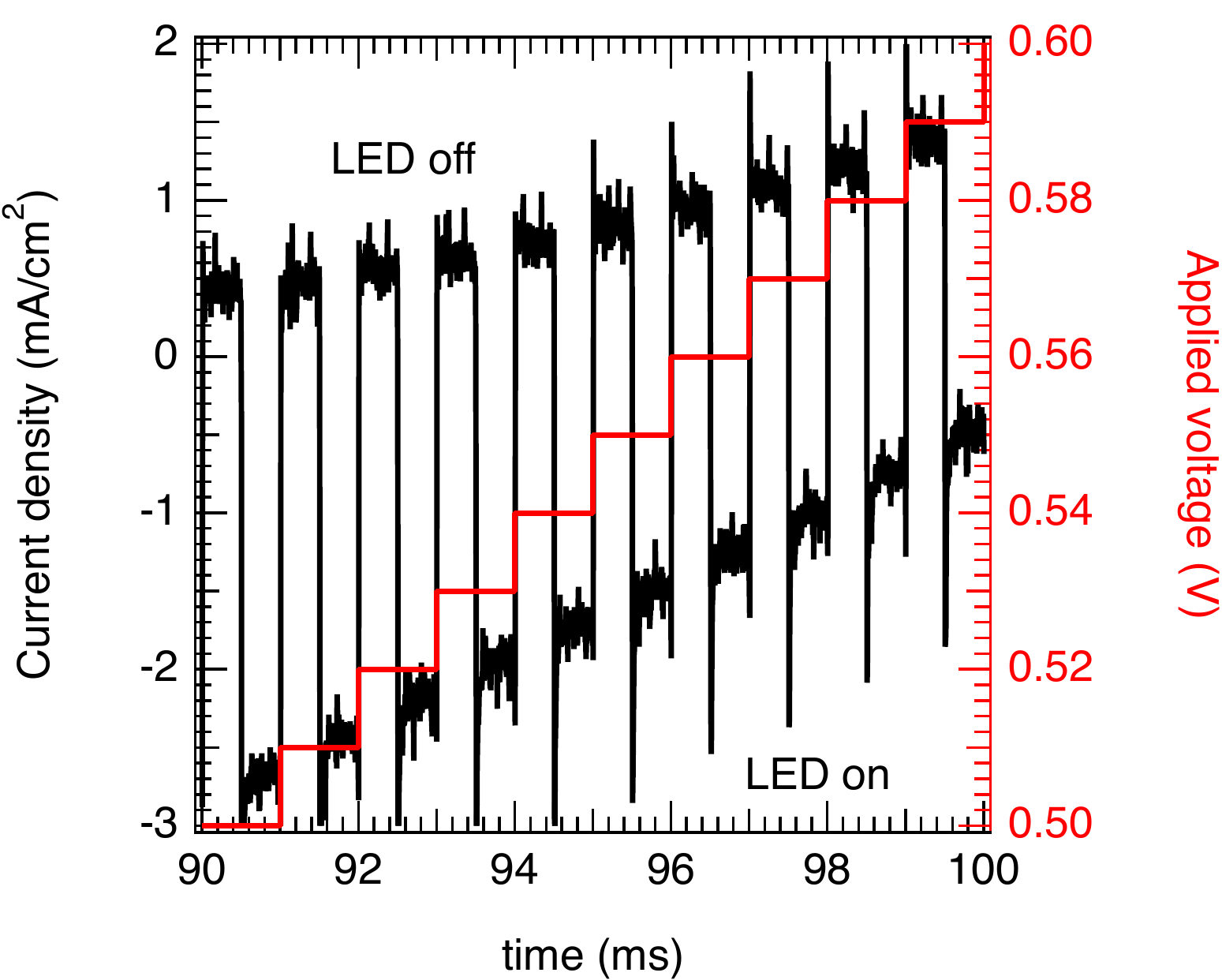}
	\caption{(Color online) Time--resolved current density (black) of a pulsed photocurrent measurement. At each voltage step (red) the current density is measured in dark and under illumination. The pulse duration was chosen sufficiently long (0.5 ms) for the current density to reach steady-state conditions after switching the LED on or off.}
	\label{fig:pulse}
\end{figure}

Pulsed photocurrent measurements were done by means of a 10 W white light emitting diode (LED) as described before. \cite{limpinsel2010, ooi2008} To provide enough time for the device to reach steady-state conditions after switching the illumination on or off, the pulse duration was set to 0.5 ms at a duty cycle of 50\% as shown in Fig.~\ref{fig:pulse}. At each voltage step (red line) the median of the steady-state values was used to calculate the resulting photocurrent density as the difference of light an dark current. 

Capacitance--voltage (CV) measurements were carried out in the dark with an Agilent E4980A LCR-meter in parallel RC-circuit mode. The validity of this operation mode was confirmed by preceding impedance measurements that showed a clear semicircle in Cole--Cole representation with a negligible offset of the real part ($\sim~50~\Omega$) which was attributed to a series resistance. For CV measurements the dc bias voltage sweep (-2 V to 2 V) was superimposed by a low frequency (5 kHz) ac voltage with a small amplitude (40 mV) to prevent any influence of the ac signal on the measured capacitance. 

The active layer thicknesses of the presented samples were in the range from 65 to 250 nm. They were determined with a Veeco Dektak 150 profilometer and verified by capacitance--voltage measurements as follows. We determined the geometric capacitance \cgeo~for high reverse bias and calculated the active layer thickness $d$ from \cgeo = $\epsilon \epsilon_{0}\frac{A}{d}$, where $A$ is the active area of the device and $\epsilon_{0}$ the dielectric constant. We used a relative permittivity of $\epsilon$ = 3.3 for a 1:1 blend ratio of P3HT:PC$_{61}$BM which was determined by our measurements (not shown) and is similar to literature values from Koster et~al. ($\epsilon$ = 3.4). \cite{Koster2006} The resulting active layer thicknesses were in good agreement with our profilometric measurements.

%\section{Results and Discussion}
%\subsection{Validity of Mott--Schottky analysis}

\begin{figure}
	\includegraphics[width=7cm]{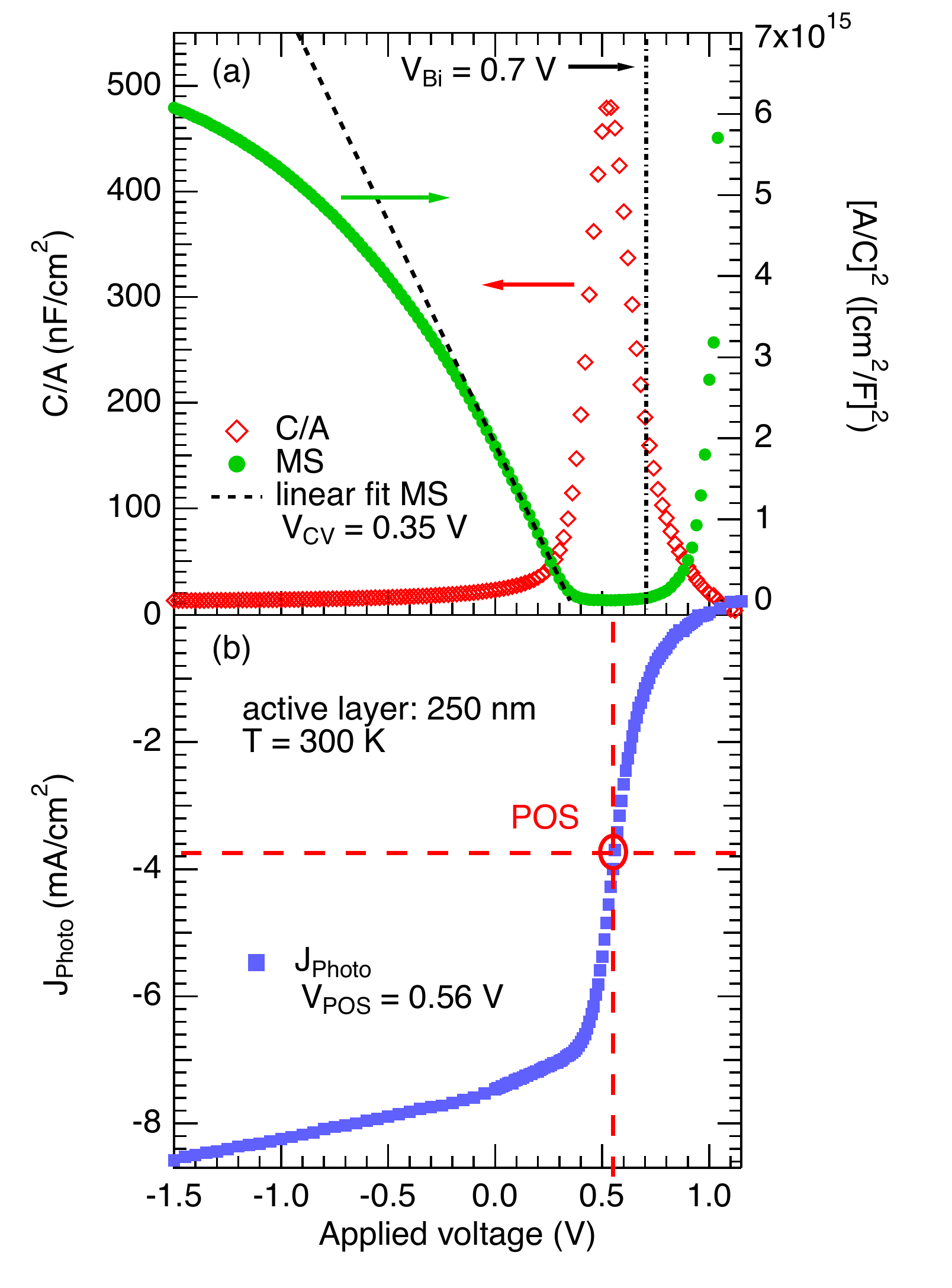}
	\caption{(Color online) Both graphs show a P3HT:PC$_{61}$BM BHJ solar cell with an active layer thickness of 250 nm at 300 K. (a) Determination of \vcv~via Mott--Schottky analysis based on capacitance--voltage measurements. (b) Determination of the point of optimal symmetry (\pos) via pulsed photocurrent measurements.}
	\label{fig:methods}
\end{figure}

In Fig.~\ref{fig:methods} (a) a typical capacitance--voltage measurement is shown (open red diamonds) in combination with MS analysis (solid green points). An interpretation of this data was proposed by Bisquert et al., \cite{bisquert2008} who distinguish three different capacitance regions with respect to the built-in potential. At reverse bias ($V$ $\ll$ \vbi) the active layer is totally depleted and therefore the measured capacitance is constant and corresponds to the geometrical capacitance (\cgeo). In the low forward bias region ($V$ $<$ \vbi) the capacitance increases inverse--quadratically (Mott--Schottky behavior). The latter is taken into account for MS analysis. The intercept of the linear fit to MS with the abscissa (dashed black line) is interpreted as \vbi~in inorganic devices. With further increasing forward bias exceeding \vbi~the capacitance decreases again due to injection. 

\begin{figure}
	\includegraphics[width=7cm]{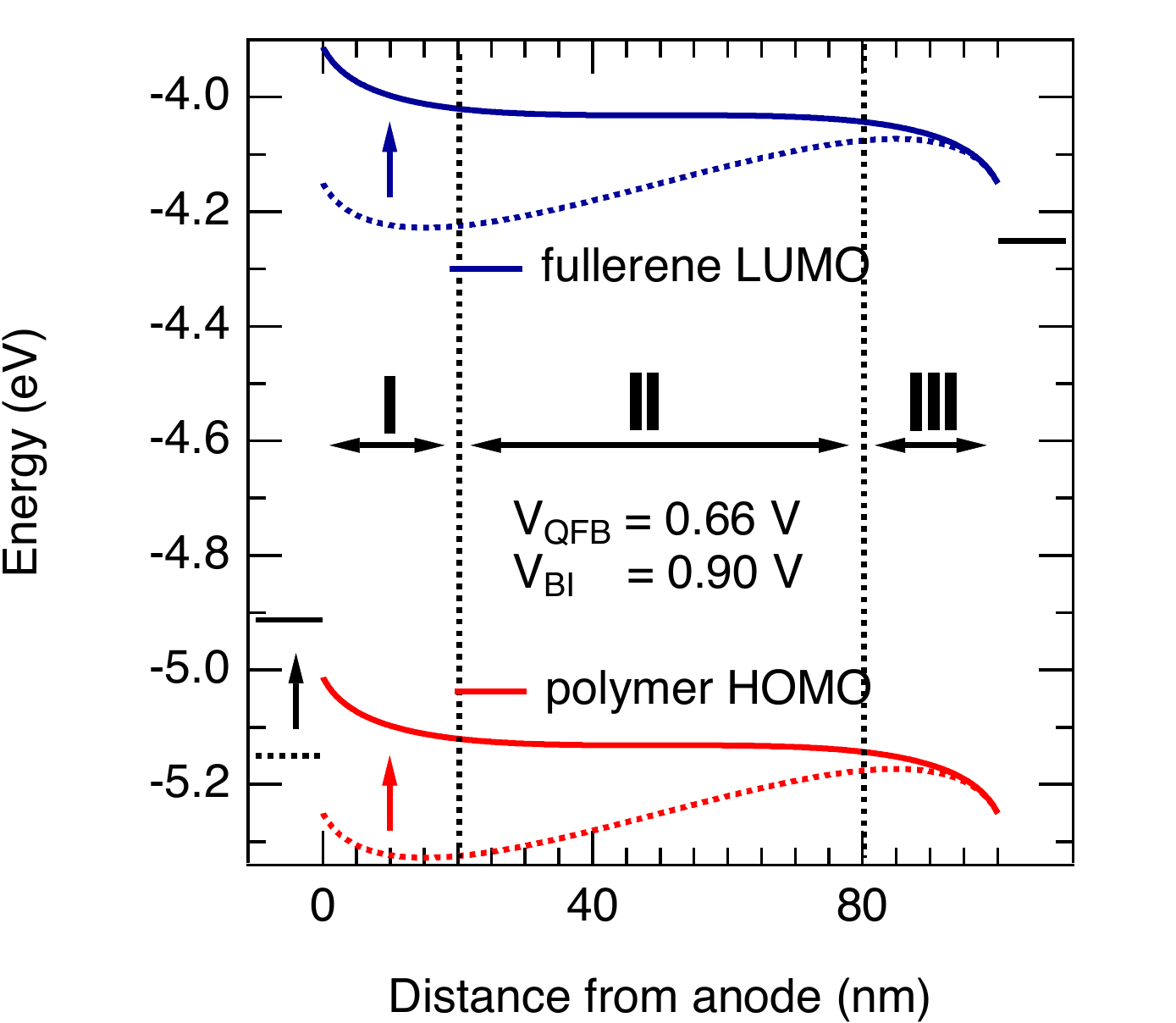}
	\caption{(Color online) Energy band diagram of an simulated organic solar cell with ITO contact (left side), metal contact (right side) and an active layer thickness of 100 nm. At an applied voltage equal to \vbi~(dashed lines) the simulated HOMO and LUMO levels match at the contacts and due to band-bending the local electric field is finite over the whole device. At a lower applied voltage \qfb~(solid lines) flat energy bands in the bulk of the device (II) are achieved, while the electric field is finite in vicinity of the contacts (I and III).}
	\label{fig:sim}
\end{figure}

We generally agree with this interpretation. Nevertheless, in pn-junctions \vbi~refers to the case of flat energy bands and therefore a zero field situation in the whole device. For organic bulk heterojunctions considerable band-bending occurs in the vicinity of metal contacts. \cite{Simmons1971} The diffusion-induced band-bending can be calculated for a semi-infinite system with zero net injection by taking the drift--diffusion equation, Einstein relation and the Poisson equation into account. \cite{Kemerink2006} These results match our simulations of the band structures, solving the one-dimensional Poisson equation and the continuity equations for electrons and holes in a numerical iterative approach. \cite{limpinsel2010} In both cases band-bending of several tenths of eV has been shown. Following these models the bands are not flat in the bulk of the device at the built-in potential, resulting in a finite electric field in the whole device due to band-bending at the contacts. Hence, a case of flat bands across the whole device does not exist (at $T~>~0~K$) which indicates that MS analysis does not exactly lead to the built-in potential (\vcv~$\neq$~\vbi). This is shown in Fig.~\ref{fig:sim}: the dashed lines of polymer HOMO (red), fullerene LUMO (blue) and ITO contact (black, left side) correspond to an applied voltage that equals~\vbi.

Instead, a case of flat energy bands in the bulk of the device (region II in Fig.~\ref{fig:sim}) can be reached at a potential well below \vbi, \cite{Kemerink2006} which we denote as quasi flat band potential \pos. \cite{limpinsel2010} Here, the resulting electric field in the bulk of the device is zero, but finite in vicinity of the contacts (region I and III) In Fig.~\ref{fig:sim} the solid lines correspond to the case of quasi flat bands. The energy level of the ITO contact (black, left side) was shifted from~\vbi~to match~\qfb.  

Another problematic issue is that MS theory describes an abrupt pn-junction and can also be applied for a Schottky contact. \cite{Sze2007} Both cases are based on bilayer systems. But organic BHJ solar cells consist of a blend of two materials and therefore we have to consider an effective medium. In this case they have to be interpreted as an ambipolar device with two slightly non-ohmic contacts, which does not correspond to MS theory.

%\subsection{Thickness Dependence}

\begin{figure}
	\includegraphics[width=7cm]{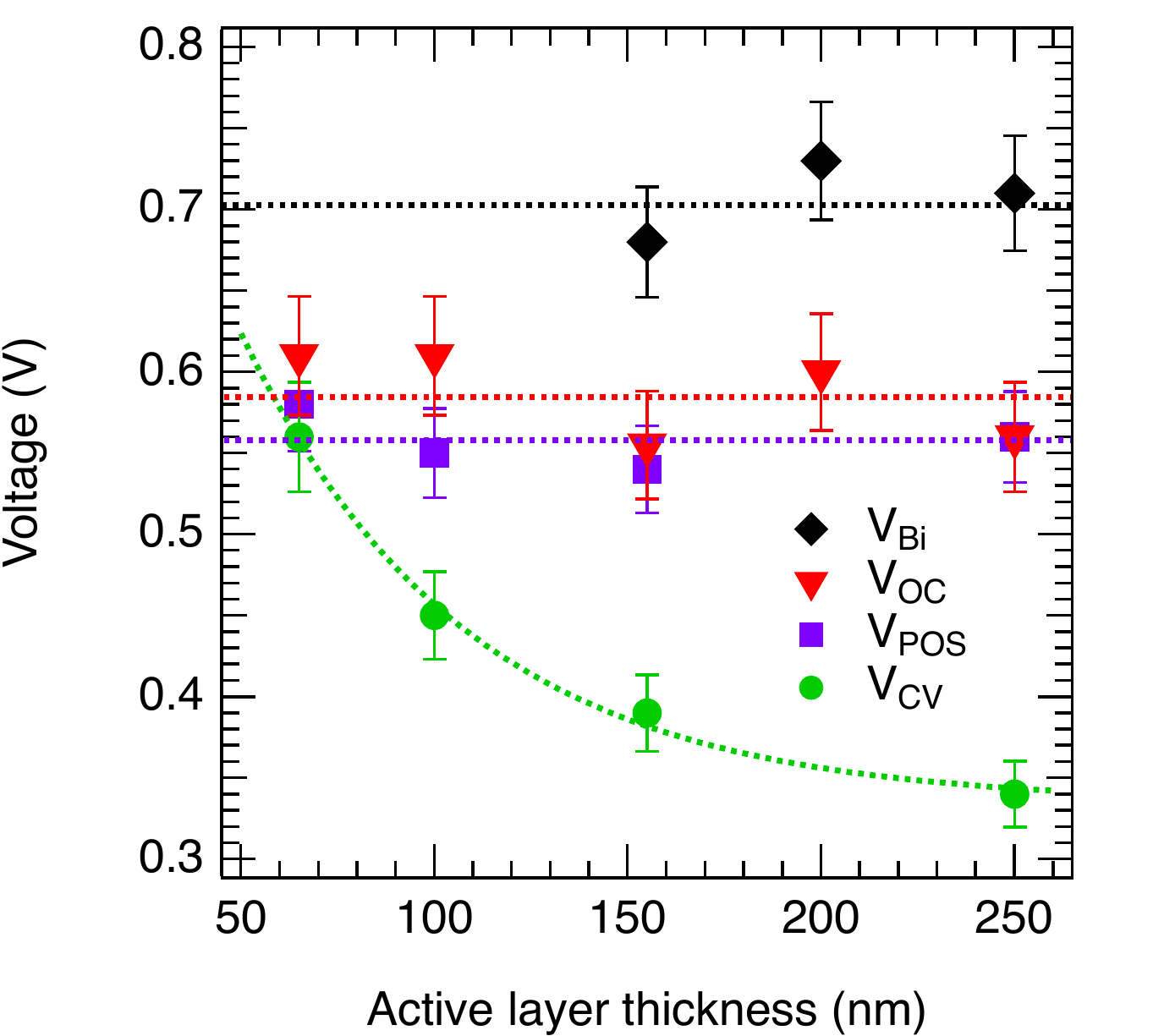}
	\caption{(Color online) \vbi~(black diamonds, determined as shown in Fig.~\ref{fig:temp}), \voc~(red triangles), \pos~(purple squares) and \vcv~(green circles) for P3HT:PC$_{61}$BM bulk heterojunction solar cells with varying active layer thickness. In contrast to \voc, \pos~and \vbi,~\vcv~shows a clear dependency on the active layer thickness (increases with decreasing thickness) and approaches \pos. The dotted lines are guides to they eyes and mark the mean values of the corresponding potentials resp. an exponential fit to \vcv.}
	\label{fig:thickness}
\end{figure}

The experimentally determined point of optimal symmetry (POS; see Fig.~\ref{fig:methods} (b)) denotes the reference for the point-symmetry of the photocurrent and corresponds to QFB. \cite{limpinsel2010} As shown in Fig.~\ref{fig:thickness} (purple squares) for P3HT:PC$_{61}$BM it is independent of the active layer thickness and situated at about 0.56 V for all samples presented which concurs to the results presented before. \cite{ooi2008, limpinsel2010} 

Furthermore in Fig.~\ref{fig:thickness} the dependency of \vcv~(green circles), \voc~(red triangles) and \vbi~(black diamonds) on the active layer thickness is also shown. The latter has been determined by temperature dependent measurements of \voc~\cite{Rauh2011} as will be explained later on. As expected \vbi~($\sim$~0.7 V) and \voc~($\sim$~0.59 V) do not show a dependence on the active layer thickness in the observed range from 50 to 250 nm. In contrast, \vcv~increases with decreasing active layer thickness. The dotted green line shows an exponential fit to \vcv~to highlight its strong thickness dependence. We find \vcv~in the range of 0.35 V (250 nm active layer) up to 0.56 V (65 nm active layer). Hence, for thin samples (below 100 nm active layer thickness) \vcv~may be close or even equal to \pos~as shown in our previous work, \cite{limpinsel2010} while there is a clear difference in both values for thick samples above 150 nm active layer.

Our values deduced from Mott--Schottky analysis are similar to those presented by other groups. For example Garcia-Belmonte et al. found a Mott--Schottky potential of 0.43 V for P3HT:PC$_{61}$BM with an active layer thickness of 200 nm, applying Al as top contact. \cite{garciabelmonte2008} 

The exponential increase of \vcv~with decreasing active layer thickness cannot be explained by standard MS analysis since the capacitance of a Schottky contact can be expressed as follows without any thickness dependence: \cite{Sze2007}

\begin{equation}
\frac{1}{C^2} = \frac{2}{q\epsilon_{s}NA^2} \left( V_{Bi} - V - \frac{kT}{q} \right).
\end{equation}   

Here, \vbi~is the built-in potential, $V$ the applied voltage, $A$ the active area of the device, $q$ the elementary charge, $\epsilon_{s}$ is the relative dielectric constant and $N$ the effective doping density. This can be seen as a first experimental indication that MS analysis does not necessarily lead to the built-in potential in organic BHJ solar cells.

%\subsection{Temperature Dependence}

\begin{figure}
	\includegraphics[width=7cm]{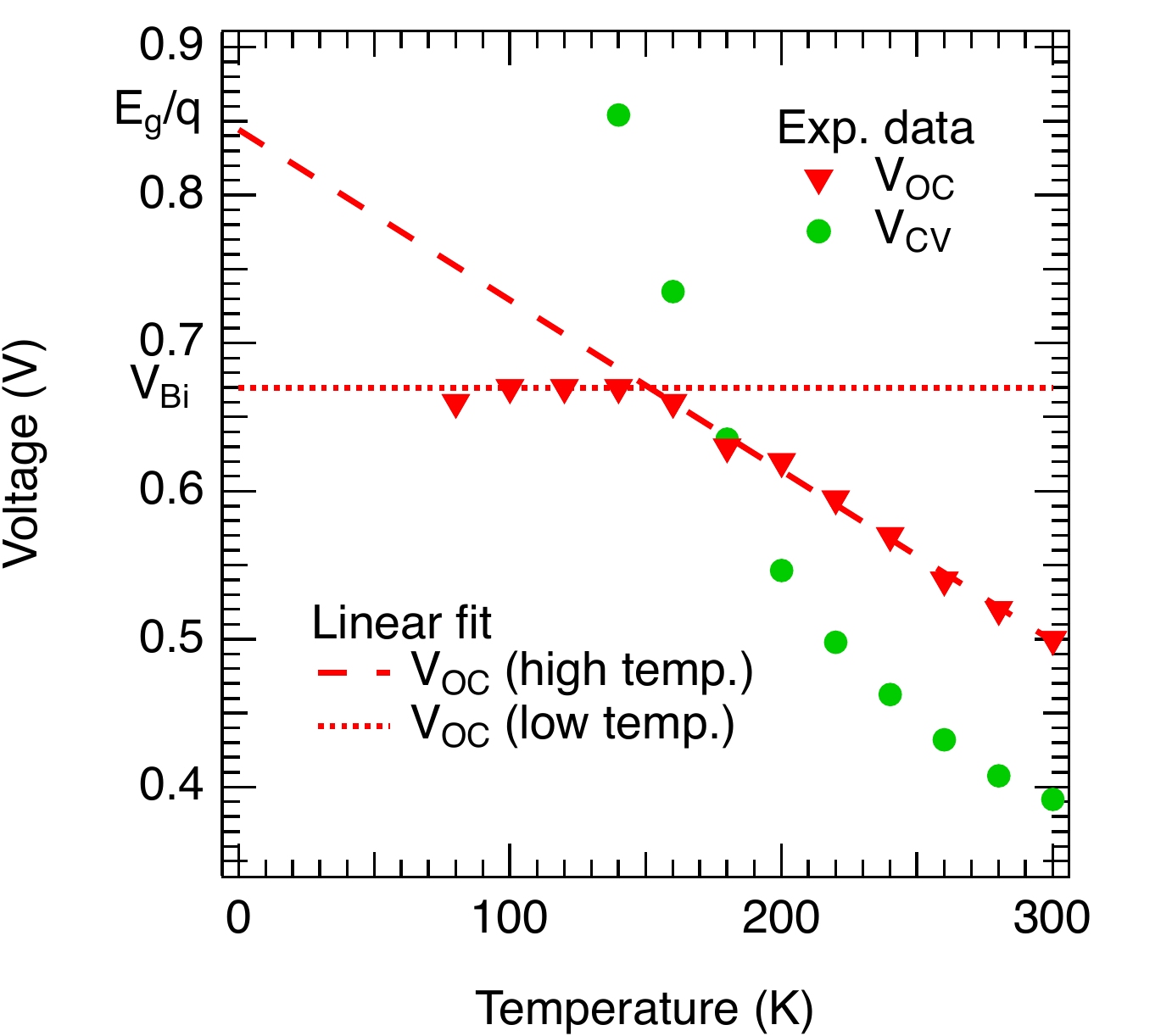}
	\caption{(Color online) P3HT:PC$_{61}$BM solar cell (active layer thickness: 155 nm). An extrapolation of the low temperature region of \voc~(dotted red line) leads to the built-in potential, while a linear extrapolation of the high temperature region (dashed red line) leads to the effective band gap $E_{g}$ (at $T$ = 0 K). \cite{Rauh2011} \vcv~(green dots) increases exponentially with decreasing temperature and shows no connection to the built-in potential.}  
	\label{fig:temp}
\end{figure}

In addition, we determined the built-in potential by temperature dependent measurements of the open circuit voltage. As shown in Fig.~\ref{fig:temp} (red triangles) \voc~increases linearly with decreasing temperature (high temperature region above 160 K), reaching a constant value (low temperature region below 160 K) when energetic barriers at the contacts dominate the behaviour of the device. We have previously shown experimentally and by simulations that the constant value for low temperatures yields \vbi, while a linear fit to the high temperature region leads to the effective band gap ($E_{g}$) at zero Kelvin. \cite{Rauh2011} The resulting built-in potentials reach values between 0.67 and 0.73 V that are not dependent on the active layer thickness (see Fig.~\ref{fig:thickness}, black diamonds). They clearly exceed the potentials derived from MS analysis (0.35 to 0.56 V, green circles) and therefore support the idea that MS analysis does not lead to the built-in voltage.

Temperature dependent capacitance--voltage measurements (see Fig.~\ref{fig:temp}, green dots) show an exponentially increasing \vcv~with decreasing temperature and with neither connection to built-in potential nor to case of quasi flat bands.

It is now possible to calculate the magnitude of diffusion-induced band-bending at both contacts. The difference between \vbi~and \pos~for P3HT:PC$_{61}$BM at room temperature is about 0.14 V which could be achieved by two injection barriers of about 0.07 eV each. These results fit well to estimations of the electron and hole injection barriers $\Phi_{e}$ and $\Phi_{h}$ of P3HT:PC$_{61}$BM by $\Phi_{e}+\Phi_{h}=E_{g}-V_{Bi}/q \approx 0.14~eV$.

%\section{Conclusions}

In conclusion, we showed that the potential obtained from MS analysis (\vcv) of organic BHJ solar cells does not correspond to the built-in potential as it is the case for inorganic pn-junctions. A comparison of \vcv~(0.35 to 0.56 V) and \vbi~($\sim$ 0.7 V) shows a clear discrepancy. In addition, \vcv~shows an unexpected dependency on the active layer thickness of the device that cannot be explained by classical MS theory. Temperature dependent measurements do not show any connection to the built-in potential as well. 

In contrast we show that the case of quasi flat bands is a more reliable measure to determine flat band conditions in an organic BHJ solar cell. It was determined by pulsed photocurrent measurements and is well below the built-in potential. \pos~is independent on the active layer thickness.

\begin{acknowledgments}

The authors thank the Bundesministerium f\"{u}r Bildung und Forschung for financial support in the framework of the MOPS project Contract No. 13N9867. V.D.'s work at the ZAE Bayern is financed by the Bavarian Ministry of Economic Affairs, Infrastructure, Transport, and Technology.

\end{acknowledgments}

\end{document}